\documentclass[a4paper,11pt]{article}

\pdfoutput=1 % if your are submitting a pdflatex (i.e. if you have
\usepackage{jcappub} % for details on the use of the package, please
\usepackage{graphicx,color}
\usepackage[dvipsnames]{xcolor}
\usepackage{amsmath}
\usepackage{booktabs}
\usepackage{hyperref}

\usepackage[T1]{fontenc} % if needed
\usepackage{capt-of}
\usepackage{multirow}

\usepackage{tikz}
\usetikzlibrary{arrows.meta,positioning,calc} % needed by the figure
\usepackage{xcolor}

\usepackage[markup=underlined]{changes} % options: final (hide markup), draft (show)
\definechangesauthor[name=Rev, color=red]{rev}

\newcommand{\gr}{$\gamma$-ray }

\usepackage[normalem]{ulem}

\title{Weakly supervised machine learning for model-agnostic searches of new phenomena in the $\gamma$-ray sky}

\author[a]{Michael Kr\"amer,}
\author[b,c]{Silvia Manconi}
\author[a]{and Kathrin Nippel}

\affiliation[a]{Institute for Theoretical Particle Physics and Cosmology, RWTH Aachen University, D-52056 Aachen, Germany}
\affiliation[b]{Laboratoire d'Annecy-le-Vieux de Physique Th\'eorique (LAPTh), CNRS, USMB, F-74940 Annecy, France}
\affiliation[c]{Sorbonne Universit\'e \& Laboratoire de Physique Th\'eorique et Hautes \'Energies (LPTHE),
CNRS, 4 Place Jussieu, Paris, France}

\emailAdd{mkraemer@physik.rwth-aachen.de, manconi@lpthe.jussieu.fr, nippel@physik.rwth-aachen.de}
\arxivnumber{TTK-26-20, LAPTH-040/26}

\begin{document}

% --------------------------------------------------
% ABSTRACT
% --------------------------------------------------

\abstract{
The $\gamma$-ray sky, as observed by the Fermi Large Area Telescope, contains a significant number of unassociated sources that may point to new astrophysical populations or more exotic phenomena. Machine-learning methods are widely used for source classification and searches for new physics, but most existing approaches rely on fully supervised training and therefore on explicit signal models.
We explore weakly supervised classification as a less model-dependent strategy for analysing $\gamma$-ray source spectra. In a background--versus--mixture setup, classifiers are trained on samples with different signal admixtures rather than on fully labelled signal and background events. We study three representative scenarios: pulsar--active galactic nuclei separation as a controlled benchmark, the identification of dark-matter subhalos, and spectral irregularities induced by axion--photon oscillations. In each case we investigate the impact of signal fraction and sample composition on classification performance.
Our results show that weak supervision can identify anomalous or signal-like subsets of data while reducing the reliance on detailed signal templates during training. In favourable cases, the method approaches the performance of fully supervised classifiers, while remaining applicable in situations where the signal model is uncertain or only partially specified. Weakly supervised learning therefore provides a complementary candidate-selection and anomaly-ranking strategy for $\gamma$-ray data analysis and searches for new phenomena.
}

\maketitle

% --------------------------------------------------
% INTRODUCTION
% --------------------------------------------------

\section{Introduction}

The $\gamma$-ray sky, observed with unprecedented accuracy by the \textit{Fermi} Large Area Telescope (Fermi-LAT) \cite{2009ApJ...697.1071A} contains a variety of sources. Many of them have been firmly identified as active galactic nuclei (AGN) or pulsars, but a significant fraction of the sources listed in the LAT catalogs remain unassociated \cite{Fermi-LAT:2019yla,Ballet:2023qzs}. These unassociated objects may correspond to new astrophysical populations, or even to more speculative scenarios such as dark matter subhalos \cite{Springel:2008cc,Madau:2008fr,Ackermann:2012nb}. 
In addition to new source populations, physics beyond the Standard Model may manifest itself through distortions of otherwise smooth source spectra, for example via axion-photon oscillations in astrophysical magnetic fields, which induce characteristic spectral irregularities \cite{Hooper:2007bq,Fermi-LAT:2016nkz}. A reliable classification of $\gamma$-ray sources is also crucial for population studies and for quantifying the contribution of different source classes to the diffuse $\gamma$-ray and neutrino backgrounds.

Machine learning has become a central tool for these classification tasks. Previous studies have applied supervised methods to Fermi-LAT data, using labelled training samples from already identified sources or from simulations. This includes both the classification of known astrophysical classes, such as pulsars and AGN (see e.g.~\cite{Mirabal:2012em, SazParkinson:2016xnd, Chiaro:2016noj, Salvetti:2017nkp, Finke:2020nrx,Butter:2021mwl}), and machine-learning-based searches for exotic candidates such as dark matter subhalos (see e.g.~\cite{Mirabal:2016huj, Coronado-Blazquez:2019puc, Mirabal:2021ayb, Gammaldi:2022wwz, Amerio:2025dmshalos,Butter:2023piw}). These studies have demonstrated the power of machine learning in $\gamma$-ray astronomy, but they also share an important limitation: supervised classifiers require labelled data for the signal model. If the true exotic signal differs from the assumed training model, the sensitivity of the classifier can be significantly reduced.

To overcome this limitation, we explore weakly supervised learning, a class of machine-learning methods in which classifiers are trained on data with incomplete or group-level  labels rather than on labelled examples~\cite{HernndezGonzlez2016WeakSA, Zhou2018ABI}. Such approaches are especially useful when the signal is not fully specified a priori or when event-level labels are unavailable or ambiguous. A notable example comes from collider physics, where techniques such as ``classification without labels'' (CWoLa) and related mixture-based methods have been successfully applied to searches for generic signals of new physics at the Large Hadron Collider \cite{Metodiev:2017vrx, Collins:2019jip, Karagiorgi:2022qnh,Belis:2023mqs}. The central idea is to train on datasets that differ only in the proportion of signal and background events, rather than requiring exact labels for each event. In this background--versus--mixture setup, the classifier exploits differences in class composition under the assumption that the background component is distributed identically across the samples. This reduces the reliance on detailed signal modelling during training, while still enabling the identification of subsets of data with atypical or signal-like properties.
Recently, related weakly supervised and anomaly-detection strategies have also been explored in astrophysics, including searches for stellar streams and unusual stellar populations in \textit{Gaia} data \cite{Shih:2021kbt,Shih:2023jfv,Pettee:2023zra, Hallin:2025wyc}, and the characterization of $\gamma$-ray bursts from blazars \cite{Cerruti:2024dzp}. 

In this work, we apply weakly supervised classification to  $\gamma$-ray source spectra. To the best of our knowledge, weakly supervised methods have not previously been applied to $\gamma$-ray source classification in this way. Our goal is to develop a less model-dependent strategy for identifying potentially exotic or anomalous $\gamma$-ray sources, reducing the reliance on specific simulations and assumptions during training.

In the astrophysical context considered here, we do not primarily envision weak supervision as a replacement for dedicated likelihood-based searches or as a direct tool for maximizing discovery significance in a counting experiment. Instead, the method is best interpreted as a flexible candidate-selection and anomaly-ranking strategy. By training on identified astrophysical populations and unlabeled or mixed source samples, weak supervision can identify subsets of sources with atypical spectral properties while reducing the reliance on explicit signal templates during training. 
Such candidate subsets can subsequently be studied with dedicated model-dependent analyses, multiwavelength observations, or follow-up measurements. In this sense, weak supervision provides a complementary discovery-oriented framework that may help prioritize interesting targets in large $\gamma$-ray catalogs, especially in situations where the underlying signal model is uncertain or only partially specified. The present work focuses on this idealized same-background setting in order to establish the achievable behaviour of weak supervision under controlled conditions, before addressing realistic selection effects and domain shifts in future studies.

We illustrate the method on three representative cases: pulsar-AGN separation, the identification of dark matter subhalos, and the search for oscillatory spectral features from axion-photon mixing in Galactic magnetic fields.
For each case we compare the weakly supervised results to a supervised benchmark and discuss strengths, limitations, and prospects. We first validate the approach on the pulsar-AGN separation task, where performance can approach that of a fully supervised reference. We then consider a more challenging, model-driven example with dark-matter subhalos, highlighting how the signal fraction and background-mixture ratio shape classification performance. Finally, we extend the method to spectral oscillations expected from axion-photon mixing in Galactic magnetic fields and outline how it may be applied to unassociated sources under realistic selection constraints. 
By progressing from a well-controlled benchmark to increasingly subtle and model-dependent scenarios, we aim to demonstrate that weak supervision can be a flexible and complementary addition to the toolkit for analysing \gr data, while also connecting astrophysical applications to developments in machine learning and particle physics.

The paper is organized as follows. Section~\ref{sec:methods} summarises the supervised and weakly supervised classification methods used in this work. Sections~\ref{sec:pulsars}--\ref{sec:alps} present three applications in turn: the separation of pulsars from AGN, the identification of dark matter subhalos, and the search for spectral features from axion--photon mixing. Our conclusions are presented in Section~\ref{sec:conclusion}. Technical details on the generation of additional \gr spectra with normalizing flows and on parameter choices are provided in the appendices.

% --------------------------------------------------
% METHODS
% --------------------------------------------------

\section{Classification methods: supervised baseline and weak supervision}
\label{sec:methods}

In this section we describe the machine-learning methods used in our study.
We first establish a supervised benchmark, which provides a reference point to assess the performance of weakly supervised classification. We then introduce the weakly supervised approach, which relaxes the requirement of fully labelled training data and allows us to construct more model-independent classifiers. 

\subsection{Supervised classification}

To quantify the performance of weak supervision, we first construct a supervised baseline with fully labelled training data. 
The goal is to learn the relationship between the input features of the $\gamma$-ray sources and their classes, using samples taken from identified sources or from simulations. (For brevity we refer to the source properties used by the classifier as \emph{features}.)
We note that the labels for the observed sources are taken as provided by the fourth Fermi-LAT catalog, the 4FGL \cite{Ballet:2023qzs} through identification using gamma rays (e.g. through gamma ray pulsation) or association with source emissions in other catalogs. Simulations of additional sources are based on the properties derived with observed, labelled sources in the 4FGL.

We compared two standard classifiers: feed-forward neural networks and boosted decision trees (BDTs)~\cite{Friedman2001GreedyFA}. Neural networks are flexible and widely used in $\gamma$-ray astronomy. In our tests, however, BDTs showed more stable performance and slightly better separation power, and they were easier to train and optimise. We employ  the \texttt{GradientBoostingClassifier} as available in \texttt{sklearn}.

An important practical advantage of BDTs is their robustness to noisy, correlated, or partially redundant input features. This is relevant for our study since we work directly with low-level quantities derived from the measured source spectra, rather than using carefully engineered high-level variables. Our setup also allows us to include additional derived features, even if they are correlated. The stability of BDTs in such situations motivates their use as our default classifier for both the supervised baseline and the weakly supervised analyses. This robustness has been widely exploited in particle physics (see, e.g.,~\cite{Finke:2023ltw}) and is well documented in the machine-learning literature~\cite{Grinsztajn2022WhyDT}.

The BDTs are trained using gradient boosting~\cite{Friedman2001GreedyFA} 
with decision trees of maximum depth three and ensembles of up to 200 trees. 
The training objective is the cross-entropy loss. 
Hyperparameters, including the number of trees, tree depth, and learning rate, 
are selected by a simple grid search based on validation performance. 

In earlier work~\cite{Butter:2023piw,Butter:2021mwl} we also employed Bayesian neural networks, which provide uncertainty estimates for the classification. While attractive from a statistical perspective, they are computationally more demanding and are not required for the present study, where our focus is the comparison between supervised and weakly supervised strategies on equal footing. 

\subsection{Weakly supervised classification}\label{subsec:weak}

Supervised classifiers require labelled data for all classes. In searches for new $\gamma$-ray sources this is limiting, because the properties of potential exotic sources are not fully known. Weakly supervised methods address this by training on \emph{mixtures} of source populations that differ only in the relative abundance of the component of interest. Individual labels are not required, yet the classifier can still learn a useful separation.

Formally, let $p_E(x)$ and $p_A(x)$ denote the feature distributions of exotic and astrophysical sources, respectively.
Two mixed samples,
\begin{equation}
p_{M_i}(x) \;=\; f_i\,p_E(x) + \bigl(1-f_i\bigr)\,p_A(x)\,, \qquad i\in\{1,2\}\,,
\end{equation}
where $f_i \in [0,1]$ denotes the fraction of exotic sources in sample $M_i$, are sufficient: if $f_1 \neq f_2$, training a classifier to distinguish $M_1$ from $M_2$ implicitly learns a decision boundary correlated with the separation between exotic and astrophysical sources, even without pure labelled examples~\cite{Metodiev:2017vrx}. Related weak-supervision paradigms and practical extensions have been explored in collider physics, including learning from label proportions and systematic studies of mixed-sample training~\cite{Collins:2019jip}, as well as approaches that combine mixed-sample classification with explicit background modelling using density estimators such as normalizing flows~\cite{Hallin:2021wme}.

In our $\gamma$-ray application, well-identified astrophysical sources such as AGN or pulsars provide reliable background examples. This allows a simplified ``background--versus--mixture'' (BvM) setup: we take a background sample $B$ with $B\sim p_A(x)$ and a mixed sample $M$ with $M\sim f\,p_E(x)+(1-f)\,p_A(x)$. Under the same-background assumption (the $p_A$ component is distributed identically in $B$ and $M$), a classifier trained to separate $B$ from $M$ learns a score that is monotone in the optimal exotic--versus--astrophysical likelihood ratio $p_E(x)/p_A(x)$, since $p_M(x)/p_B(x) = f\,p_E(x)/p_A(x) + (1-f)$ is strictly increasing in $p_E(x)/p_A(x)$ for $f\in(0,1]$. Thus, in the limit of sufficient training data and an expressive classifier, thresholds on the classifier score correspond to increasing evidence for an exotic origin, without requiring individual labels or prior knowledge of the signal fraction. We will use this BvM variant in sections~\ref{sec:pulsars}--\ref{sec:alps}.
We note that the notion of background can vary among different, specific applications: while in collider physics it usually refers to the signals produced by particles in the Standard Model, in our $\gamma$-ray application the ''background'' source class is meant to be the dominant, astrophysical class type.

As in the supervised case, we employ BDTs as our default classifier. They provide stable training and are robust against potential noise or redundancy in the inputs, which is important when mixtures are constructed from different source populations.

Finally, we note several practical caveats. Weak supervision assumes that the samples differ \emph{only} in class proportions; any additional distributional differences (e.g.\ selection effects or correlations between mixture-definition variables and inputs) can bias the result. Finite sample sizes and class imbalance also affect performance, and separation is intrinsically difficult if exotic spectra closely resemble astrophysical ones. We therefore validate the approach on controlled test cases and compare against a supervised benchmark trained on labelled data.

% --------------------------------------------------
% APPLICATION 1
% --------------------------------------------------

\section{Pulsars vs AGN: a controlled benchmark}\label{sec:pulsars}

In this section we study the classification of pulsars versus AGN as a controlled benchmark for weakly supervised learning. Both source classes are well established, a large number of labelled examples is available, and supervised classifiers are known to achieve high performance using spectral information alone~\cite{Finke:2020nrx,Butter:2021mwl}. This setting provides a clean reference case in which the supervised solution is well understood, allowing a direct comparison with weakly supervised strategies.
Throughout this section, we treat pulsars as the positive class and AGN as background. We first describe the data sample, then present the supervised benchmark, and finally compare it with the weakly supervised approach.

\subsection{Physics motivation and data sample}

As a common data basis we use the 4FGL Fermi-LAT catalog (DR4, 14-year release~\cite{Ballet:2023qzs}, \texttt{gll\_psc\_v32.fit}). It contains 4006 sources classified as AGN, 320 pulsars, 2430 unassociated sources (UNID), and 439 additional identified sources belonging to other astrophysical classes.
AGN and pulsars represent the two most numerous identified source classes in the catalog and exhibit distinct spectral characteristics, which makes them a natural and well-controlled test case. In this section we restrict ourselves to sources with firm associations. The unassociated sources are not used here, but we note that they constitute a natural target for future follow-up applications.

\subsection{Simulation and feature representation}\label{subsec:agnpsr_sim}
For the supervised benchmark discussed in section~\ref{subsec:agnpsr_supervised}, no simulated data are required, since both AGN and pulsars are available as labelled classes in the 4FGL catalog. Simulations enter only in the weakly supervised setup, where sufficiently large background samples are needed to probe small signal fractions and realistic class imbalance.

Throughout this section, the baseline input representation consists of the \gr fluxes in the seven standard energy bands provided in the 4FGL catalog (column labelled \texttt{Flux\_Band}), covering the energy range from 50~MeV to 1~TeV. These seven flux values define a seven-dimensional feature space encoding the coarse spectral shape of each source. Pulsars typically exhibit curved spectra with exponential cutoffs, while AGN are often characterised by power-law-like spectra. Previous studies~\cite{Finke:2020nrx,Butter:2021mwl} have shown that this spectral information alone already provides significant discriminating power between the two classes.

We therefore adopt the flux bands as low-level spectral features in a model-agnostic baseline setup. Additional catalog-level quantities such as time variability indicators or positional information are not included in this baseline configuration, in order to avoid relying on source-class-specific information and to ensure a controlled comparison between supervised and weakly supervised strategies. The impact of extending the input space with further derived features will be investigated separately below.

To construct sufficiently large background samples for the weakly supervised analysis, we generate additional AGN $\gamma$-ray spectra using a normalizing flow trained on the observed AGN population in the 4FGL catalog. Normalizing flows provide a flexible and fully invertible generative model capable of learning complex, multimodal distributions while allowing efficient sampling~\cite{JimenezRezende2015VariationalIW,Dinh2016DensityEU}. In our case, the dimensionality of the data is modest (seven spectral energy bins), which allows flow-based models to be trained efficiently and to capture the multimodal structure of the AGN distribution without requiring excessively large training samples.
We employ a normalizing flow with spline-based coupling transformations~\cite{Durkan2019NeuralSF}, which we find to be effective in reproducing the multimodal structure of AGN spectra. The flow is trained on preprocessed AGN flux spectra, where we standardize the logarithm of the fluxes in each energy bin. Once trained, the model is used to generate synthetic AGN spectra that closely match the observed per-bin flux distributions and correlations. Detailed information on the flow architecture, training procedure, and validation tests is provided in appendix~\ref{sec:app_NF}.

The generated AGN spectra are used exclusively to augment the background sample in the weakly supervised analysis. In particular, they allow us to construct background and mixed samples of controlled size and composition, enabling a systematic study of weak-supervision performance as a function of the signal fraction while preserving the characteristic spectral features of astrophysical AGN.

\subsection{Supervised classification benchmark}\label{subsec:agnpsr_supervised}

We first construct a supervised reference classifier using the seven flux-band features described above. To assess whether additional catalog-level information improves performance, we compare this low-level representation with alternative inputs available within the 4FGL catalog: derived spectral-fit parameters, the variability index (\texttt{Variability\_Index}), sky position (\texttt{GLON}, \texttt{GLAT}), flux  as a function of time (\texttt{Flux\_History}), and combinations thereof. All inputs are preprocessed consistently: quantities spanning several orders of magnitude are log-scaled, and all features are standardized.

Across inputs we obtain high accuracy (\(\sim 97~\%\)), but accuracy is not informative in the presence of significant class imbalance (4006 AGN vs 320 pulsars). More relevant are the true positive rate (TPR = TP/(TP+FN)) and the false positive rate (FPR = FP/(FP+TN)), with pulsars treated as the positive class. All quoted TPR and FPR values correspond to a fixed classifier score threshold of 0.5. We verified that the qualitative behaviour remains unchanged for other reasonable threshold choices. We find a generally low FPR in all tests. Using all numerical catalog features gives a TPR of 0.83, compared to 0.66 for the flux band alone, but this gain is almost entirely due to time variability: a BDT feature-importance analysis based on the average split-gain across the boosted trees assigns \(\sim 35~\%\) relative importance to \texttt{Variability\_Index}, versus \(\lesssim 20~\%\) for other features. Since variability is very specific to the pulsar-AGN separation and would bias a model-agnostic search, we remove it. After doing so, we conclude that the flux band is the most suitable low-level input for a model-independent setup. Adding positional or flux-history information does not further improve performance for BDTs, so we do not consider these inputs.

For the supervised benchmark we therefore use the flux band as input and BDTs as the classifier. The data are randomly split into disjoint training and test samples, preserving the class proportions. All quoted performance metrics are evaluated on data not used during training. To assess stability, the training is repeated for multiple random splits, and the reported values correspond to the mean performance; variations between runs are small compared to the quoted digits. In this configuration we obtain an accuracy of 98~\%, a TPR of 0.82, and an FPR of 0.01. These results define the supervised reference against which the weakly supervised approach is evaluated.

\subsection{Weakly supervised classification results}\label{subsec:agnpsr_cwola}

We adopt a background--versus--mixture variant of weak supervision (cf.\ section~\ref{subsec:weak}). AGN constitute the background class $B \sim p_A$, while the mixed sample $M$ contains both AGN and pulsars, $M \sim f\,p_E + (1-f)\,p_A$. If $B$ and $M$ share the same background distribution (i.e.\ the AGN component is identically distributed in both samples), a classifier trained to separate $B$ from $M$ learns a score that is informative for pulsar versus AGN separation without using individual labels.

There are 4006 AGN and 320 pulsars in 4FGL-DR4. In constructing finite background and mixed samples from these data, we require the background sample to be larger than the mixed sample ($N_B > N_M$), which limits how many AGN can be assigned to $M$; with all pulsars included, in this finite-data setup this implies a maximal signal fraction of about $15~\%$. This is useful context, since in model-dependent dark-matter subhalo searches signal fractions below $10~\%$ are expected (section~\ref{sec:darkmatter}).

To probe smaller fractions and emulate the class imbalance of the full catalog, we augment the AGN background with spectra generated by a normalizing flow as described in section~\ref{subsec:agnpsr_sim} and appendix~\ref{sec:app_NF}. In practice, we choose sizes such that the total number of background events (real AGN + generated AGN) matches the total number of labelled astrophysical sources (4765), and the size of $M$ (real AGN + generated AGN + pulsars) matches the number of unassociated sources (2430). This choice mirrors the imbalance between labelled and unassociated sources in the full catalog and fixes the overall sample sizes across signal fractions.

\begin{figure}[t]
    \centering
    \includegraphics[width = 0.75\textwidth]
    {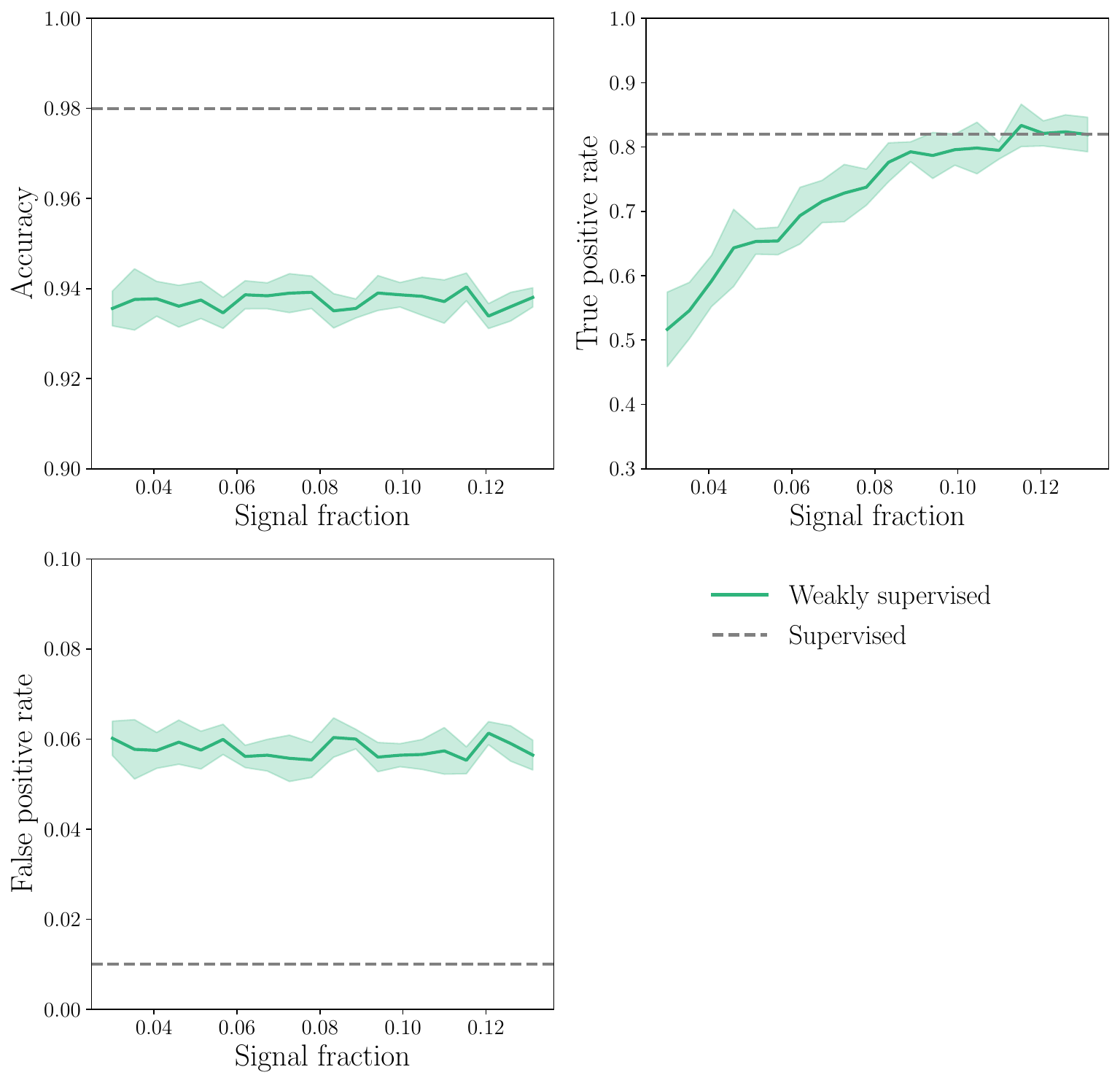}
    \caption{Accuracy, true positive rate (TPR), and false positive rate (FPR) as a function of the signal fraction for weakly supervised classification of pulsars in an AGN background. Dashed lines show the supervised benchmark for reference. }
    \label{fig:Cwola_AGN_PSR}
\end{figure}

For each probed signal fraction we train a BDT classifier multiple times with different random splits, and evaluate accuracy, TPR, and FPR using the true pulsar/AGN labels (which are not used during training) at a fixed classifier score threshold of 0.5, applied uniformly across all signal fractions (figure~\ref{fig:Cwola_AGN_PSR}). The overall accuracy remains nearly constant, which is expected given the strong class imbalance and the fact that it is dominated by the large AGN population. The FPR is also approximately stable as the signal fraction varies, while the TPR increases steadily and approaches the supervised benchmark for larger fractions.

This behaviour reflects the structure of the background--versus--mixture setup. The classifier is trained to distinguish a pure AGN sample from a mixed sample containing AGN and pulsars, rather than to separate pulsars from AGN directly. Increasing the signal fraction strengthens the pulsar contribution in the mixed sample and makes the difference between the two training samples more pronounced. This improves the separation of pulsar-like spectra and leads to the observed rise in TPR.

In contrast, the AGN distribution is identical in both training samples for all signal fractions. Increasing the signal fraction therefore does not provide additional information about the background class itself, and the FPR remains approximately constant. The overall accuracy is also largely unchanged, since it is dominated by the much more numerous AGN population and therefore closely tracks the FPR. The remaining difference with respect to the fully supervised benchmark reflects the different training objective, which results in a slightly softer decision boundary than in fully supervised training.

% --------------------------------------------------
% APPLICATION 2
% --------------------------------------------------

\section{Dark Matter Subhalos}\label{sec:darkmatter} 

Dark matter subhalos constitute a qualitatively more challenging target for classification than the pulsar-AGN benchmark discussed in the previous section. Their expected \gr spectra can closely resemble those of ordinary astrophysical sources, and searches therefore rely strongly on assumptions about the underlying particle-physics model. This makes the problem a natural test case for weakly supervised approaches that aim to reduce reliance on predefined signal labels during training while retaining sensitivity to anomalous spectra. In this section, we use simulated dark matter subhalo spectra as a controlled test signal to compare supervised and weakly supervised classification strategies on equal footing.

\subsection{Physics motivation and data sample}

The dark matter subhalo application builds on our earlier work~\cite{Butter:2023piw}, where we used Bayesian supervised feed-forward neural networks to search for dark matter subhalo candidates among 4FGL sources of unknown nature. Such subhalos could produce the observed \gr emission through dark matter annihilation or decay. In that study, the Bayesian treatment provided calibrated predictive uncertainties and, in turn, conservative limits on the dark matter annihilation cross section.

However, this type of search is highly model-dependent. In each binary classification setup, the signal model has to be predefined and simulated. For example, in \cite{Butter:2023piw} we concentrated on benchmarking weakly interacting massive particles (WIMPs) annihilating in the $b\bar{b}$ channel, as this produces energy spectra similar to those of other astrophysical sources and thus represents a challenging setup for machine-learning classifiers. If the true signal deviates from the assumed model, a purely supervised search can lose sensitivity.

In the present study, we therefore retain the same simulated subhalo spectra as a test signal to calibrate performance, but shift the focus to classification strategies that reduce reliance on fully labelled signal examples. Concretely, we first establish a supervised baseline using BDTs and the same low-level spectral inputs as in the pulsar-AGN benchmark, and then evaluate a weakly supervised background--versus--mixture setup that does not require event-level signal labels during training.

\subsection{Simulation and feature representation}\label{subsec:dm_sim}

For the dark matter subhalo study we require realistic samples of \gr spectra from subhalos for both supervised calibration and weakly supervised evaluation. To this end, we model the dark matter subhalo population using cosmological subhalo distributions derived from N-body simulations~\cite{Springel:2008cc}, which provide a statistical ensemble of subhalo positions and J-factors, i.e.\ line-of-sight integrals over the squared dark matter density that determine the expected annihilation flux.

For each benchmark dark matter model, specified by the particle mass and annihilation channel, the differential \gr energy spectrum from annihilation is computed using standard particle-physics templates~\cite{Cirelli:2010xx}. The resulting spectra are converted into expected \gr fluxes in the Fermi-LAT energy bands by integrating over energy and folding with the Fermi-LAT instrument response functions, including exposure and background models, following the procedure described in~\cite{Butter:2023piw}.

The detectability of dark matter subhalos in the Fermi-LAT catalog is then assessed by simulating the observation of these spectra under realistic conditions, including astrophysical foregrounds, isotropic background emission, and instrumental effects. Only subhalos that would be detected with sufficient significance in a likelihood analysis analogous to the 4FGL catalog construction, implemented with the standard Fermi ScienceTools and the \texttt{fermipy} framework~\cite{Wood2017FermipyAO}, are retained. Throughout this work, simulated subhalo spectra are represented using the same low-level input features as in the pulsar-AGN benchmark, namely the \gr flux in fixed energy bands. No additional derived or catalog-level features are included, ensuring a consistent and model-agnostic feature representation across supervised and weakly supervised analyses.

\subsection{Supervised classification benchmark}\label{sec:darkmatter:supervised}

We have shown in~\cite{Butter:2023piw} that supervised classification of dark matter subhalo spectra is challenging, since the predicted spectra can closely resemble those of ordinary astrophysical sources. Nevertheless, using labelled training data we achieved competitive performance and derived conservative upper limits on the dark matter annihilation cross section.

In the present study, we establish a supervised baseline with BDTs, using the same low-level spectral inputs as in section~\ref{sec:pulsars} (flux in energy bands). 
We focus on WIMP annihilation into $b\bar{b}$ for representative dark matter masses $m_\chi \in \{10, 30, 80, 300\}\,\mathrm{GeV}$ and $1\,\mathrm{TeV}$, which include cases where the dark matter spectra are particularly similar to astrophysical ones. Here the classifier is trained directly on labelled subhalo and astrophysical spectra, and thus optimises the separation between the two classes explicitly.

For illustration, at $m_\chi=80\,\mathrm{GeV}$ the supervised BDT achieves an accuracy of 94~\%, a TPR of 83~\%, and an FPR of 2~\%. Across the mass grid, the supervised results obtained here are consistent with~\cite{Butter:2023piw} for the same inputs and operating point. These supervised numbers provide the baseline against which we compare the weakly supervised results in section~\ref{sec:darkmatter:ws}. 

For completeness, we note that the data are split into statistically independent training, validation, and test samples, preserving class proportions. All quoted performance metrics are evaluated on the test set. To assess robustness, the training is repeated for multiple random splits, and the reported values correspond to mean performance; variations between runs are small compared to the quoted digits.

\subsection{Weakly supervised classification results}\label{sec:darkmatter:ws}

For the weakly supervised search for dark matter subhalos we follow the background-versus -mixture  setup introduced in section~\ref{sec:methods}. All 4FGL sources identified or associated to astrophysical (AGN, pulsars, and other classes) are used as background. Simulated subhalo spectra from~\cite{Butter:2023piw} are injected into a disjoint subset of this background pool to form the mixed sample. Background events in $B$ and $M$ are statistically independent and share the same distribution, satisfying the same-background assumption.

The weakly supervised classifier is trained to separate $B$ from $M$. As discussed in section~\ref{subsec:agnpsr_cwola}, this procedure yields a score that is informative for subhalo versus astrophysical separation without using individual signal labels.

We consider two regimes for the relative size of background and mixture: a \emph{balanced} case ($|B|{:}|M|=1{:}1$) and an \emph{imbalanced} case ($|B|{:}|M|=2{:}1$). The imbalanced setup gives more statistical weight to background events and therefore leads to a more conservative decision boundary. All performance metrics are evaluated on a held-out test set at a fixed score threshold of 0.5.

\begin{figure}
    \centering
    \includegraphics[width = 0.75\textwidth]%{Figures/Final/cwola_80GeV_grid.pdf}
    {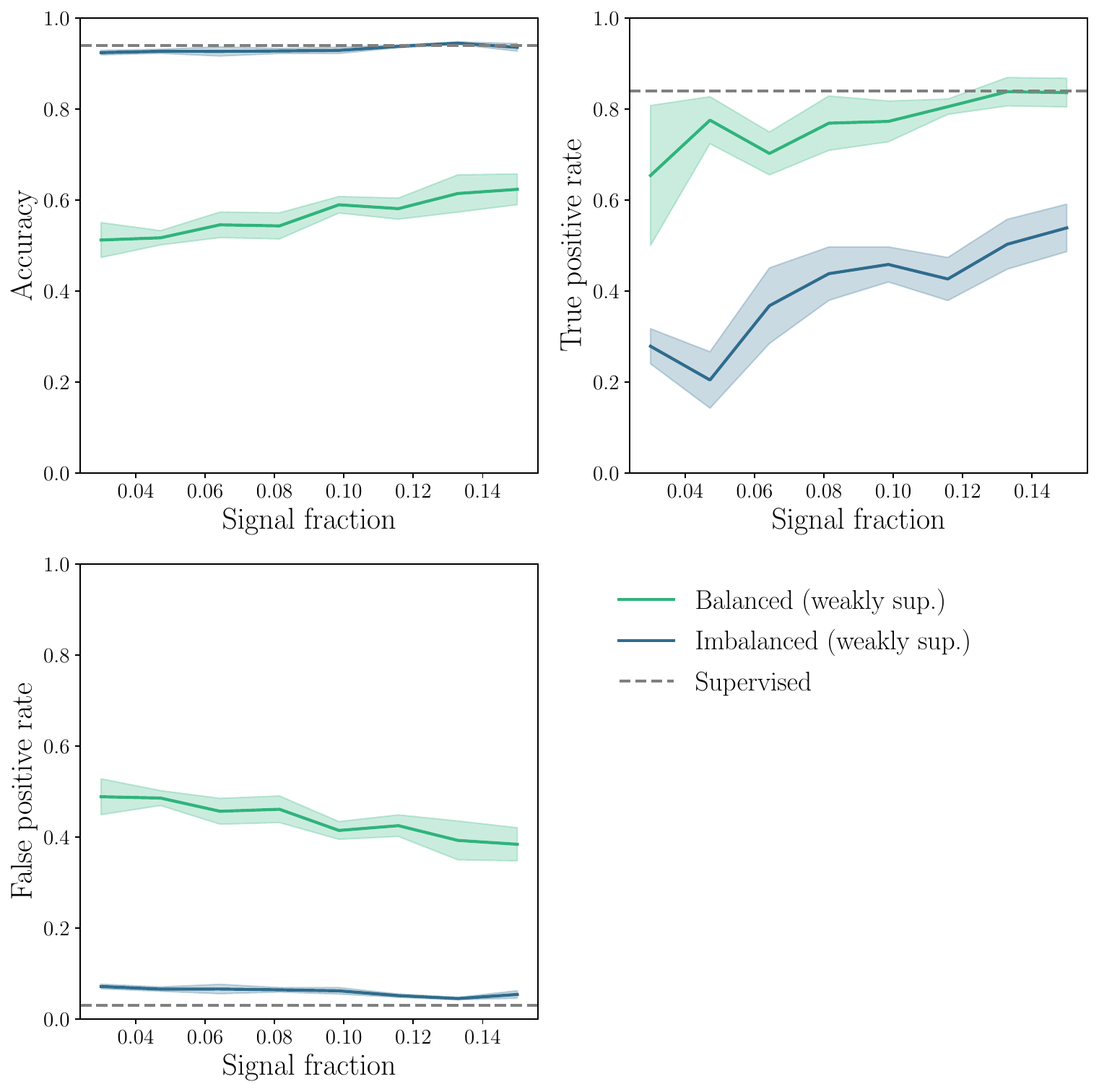}
    \caption{Analogous to figure~\ref{fig:Cwola_AGN_PSR}, but for dark matter with $m_\chi=80\,\mathrm{GeV}$ annihilating into $b\bar b$ against identified and associated 4FGL sources. Curves show accuracy, TPR, and FPR versus the signal fraction $f$ in the mixed sample. Balanced (green) and imbalanced (blue) regimes are shown. Dashed lines indicate the supervised benchmark.}
    \label{fig:Cwola_80GeV}
\end{figure}

Figure~\ref{fig:Cwola_80GeV} shows accuracy, TPR, and FPR for $m_\chi=80\,\mathrm{GeV}$ annihilating into $b\bar b$, as a function of the signal fraction $f$ in the mixed sample. In the balanced regime, the TPR increases with $f$ and approaches the supervised benchmark for larger fractions, but this comes at the cost of a significant increase in FPR. In the imbalanced regime, FPR remains nearly constant as $f$ increases, while TPR stays below the supervised baseline. The overall accuracy largely follows the behaviour of the FPR because the data set is dominated by astrophysical background sources.

These trends are consistent with the structural properties of the BvM setup discussed in section~\ref{subsec:agnpsr_cwola}. Increasing $f$ strengthens the distortion between background and mixture and improves the identification of signal-like spectra. However, in contrast to the pulsar-AGN benchmark, the background here is heterogeneous and overlaps substantially with subhalo spectra. As a result, the trade-off between TPR and FPR is more pronounced, and the supervised benchmark is not fully recovered even at larger $f$.

\begin{figure}
    \centering
    \includegraphics[width = 0.75\textwidth]%
    {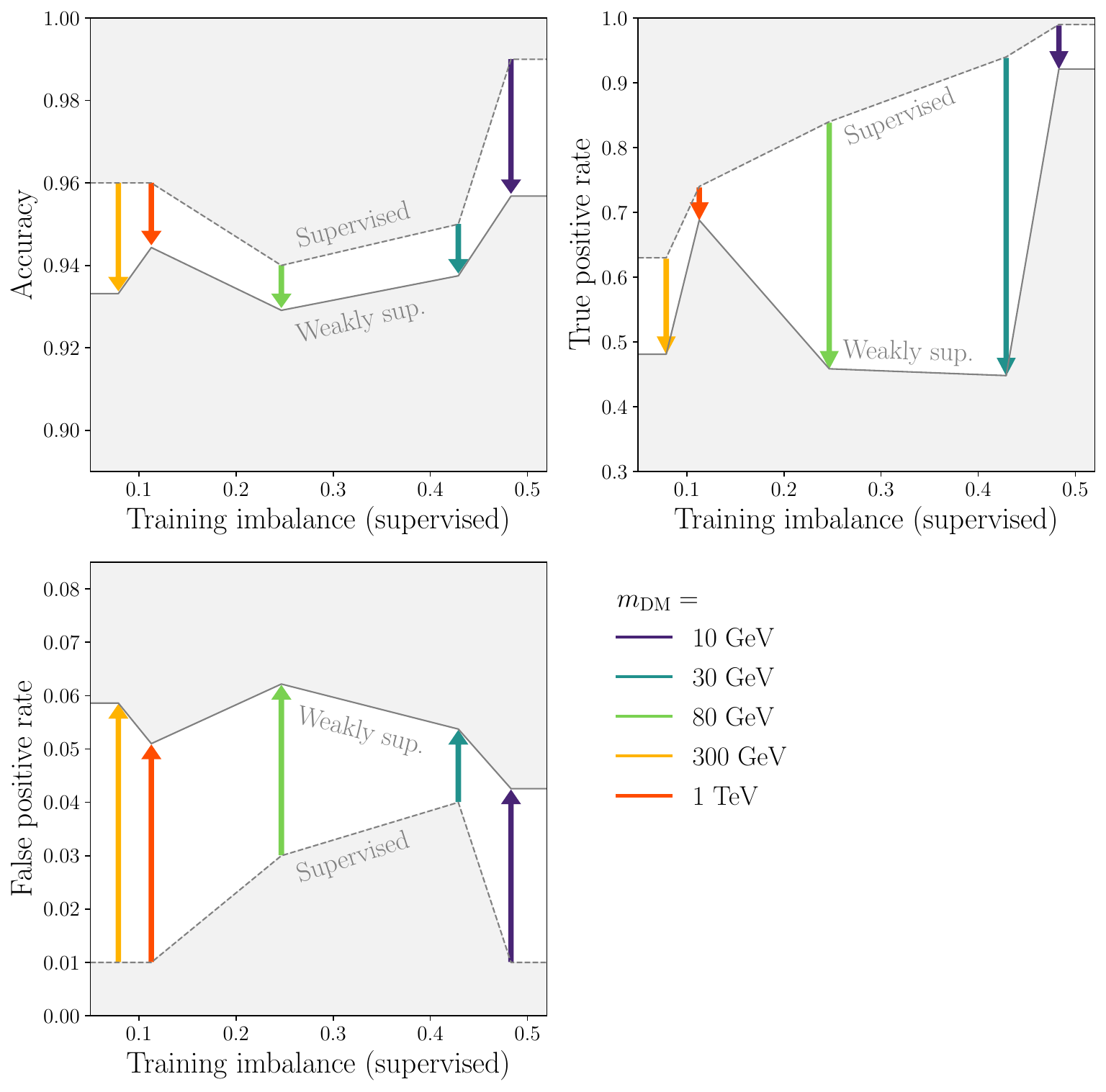}
    \caption{Comparison among the performance of supervised vs.\ weakly supervised learning on the dark matter subhalo spectra. The panels report accuracy, TPR, and FPR on a held-out test set for $m_\chi\in\{10,30,80,300\}\,\mathrm{GeV},\,1\,\mathrm{TeV}$, each indicated with different colors.
    In each panel, the results of the supervised learning correspond to the values at the tail of the arrows, while the arrow tips indicate the increased/decreased value when using the weakly supervised approach for that specific mass. Note that the x axis, indicating the training imbalance, refers only to the supervised results. The weakly supervised results are obtained  at fixed mixed-sample signal fraction $f=10~\%$ and imbalanced background-to-mixture ratio $|B|{:}|M|=2{:}1$.
    }
    \label{fig:Cwola_DM}
\end{figure}

Having established the behaviour for $m_\chi=80\,\mathrm{GeV}$, we now compare supervised and weakly supervised performance across different dark matter masses at a fixed mixed-sample signal fraction of $f=10~\%$. The results are shown in figure~\ref{fig:Cwola_DM}.

For each mass we obtain one weakly supervised result, trained in the imbalanced BvM setup with $f=10~\%$. The supervised classifiers, in contrast, are trained with labelled data sets containing a chosen fraction of simulated subhalo spectra. The x-axis in figure~\ref{fig:Cwola_DM} exactly shows this fraction, which is used during supervised training. Note that the weakly supervised point does not move along the x-axis; the imbalance corresponds to the fixed BvM setup described above for all the dark matter masses.

Each arrow connects the supervised result (arrow tail) to the corresponding weakly supervised result (arrow tip) for the same dark matter mass. Upward or rightward arrows indicates higher accuracy or TPR, while downward arrows indicates lower FPR.

Weak supervision reduces accuracy only slightly for most masses. The decrease in TPR is strongest for $30$ and $80\,\mathrm{GeV}$, where signal and background spectra are most similar, and milder for $10\,\mathrm{GeV}$, $300\,\mathrm{GeV}$, and $1\,\mathrm{TeV}$. The FPR increases modestly overall. These observations are consistent with the $80\,\mathrm{GeV}$ scan in figure~\ref{fig:Cwola_80GeV} and reflect the conservative decision boundary of the BvM setup at small signal fraction.

Overall, we find that weakly supervised classification can identify subhalo-like spectra without requiring explicit signal labels during training. The remaining performance gap relative to supervised training reflects the different optimisation objective and the substantial spectral overlap between subhalos and astrophysical sources. The dark matter problem is therefore intrinsically more challenging than the pulsar-AGN benchmark discussed in section~\ref{sec:pulsars}.

Improved separation may be expected for annihilation channels with harder spectra, for example $\tau^+\tau^-$. In such cases the similarity between signal and background spectra is reduced. Importantly, the weakly supervised setup does not require retraining for each signal hypothesis and therefore reduces reliance on detailed signal modelling.

A natural next step is to apply this approach directly to real data, using identified and associated astrophysical sources as $B$ and unassociated sources as $M$. Such an analysis requires careful control of selection effects (exposure, latitude, flux thresholds and detection-significance requirements) and validation of the same-background assumption. We leave this to future work.

% --------------------------------------------------
% APPLICATION 3
% --------------------------------------------------

\section{Spectral irregularities from ALP-photon oscillations}\label{sec:alps}

In this section we consider a qualitatively different signal scenario: spectral irregularities induced by axion-photon oscillations. Unlike the previous two case studies, the signal does not correspond to a new source population, but to subtle deformations of otherwise smooth spectra. In analogy to sections~\ref{sec:pulsars} and~\ref{sec:darkmatter}, we first describe the physical motivation and data setup, then outline the simulation of modulated spectra, and finally compare supervised and weakly supervised classification strategies.

\subsection{Physics motivation and data sample}

In the last application of the weakly supervised approach explored in this work, we search for a different type of exotic signal accessible through the inspection of \gr spectra. Apart from GeV-scale WIMP dark matter, other new types of fundamental interactions and particles can leave imprints on the spectra of cosmic accelerators as well as in cosmic backgrounds, such as the QCD axion, which provides an appealing solution to the unexplained lack of CP violation in strong interactions~\cite{DiLuzio:2020wdo}. More broadly, axion-like particles (ALPs) are often predicted in beyond the Standard Model theories, and could act as dark matter candidates in specific regions of the currently allowed parameter space, spanned by the axion mass $m_a$ and the coupling with photons $g_{a\gamma}$ \cite{Chadha-Day:2021szb,OHare:2024nmr}. 

ALPs can be converted into photons in the presence of magnetic fields. This can lead to an energy-dependent modulation of the \gr fluxes from Galactic and extragalactic sources, which effectively produces irregularities, so-called wiggles in their spectra \cite{Hooper:2007bq}. 
%see pag 13 https://arxiv.org/pdf/2505.12905 for main references 
Spectral modulations in the \gr spectra might be observable if the magnetic field is sufficiently strong and/or extended in the line of sight between the source and our telescopes, and the ALP-photon coupling is sufficiently large, see e.g.\ \cite{Fermi-LAT:2016nkz} for searches of such a signal within the intra-cluster magnetic field of NGC 1275. 
Searches have also been performed using the observed spectrum of Galactic cosmic ray accelerators such as pulsars and supernovae, assuming that the ALP-photon mixing occurs in the Galactic magnetic field \cite{Majumdar_2018,Xia:2018xbt,Liang:2018mqm,Pallathadka:2020vwu}.
In this context, hints for unexpected spectral modulations in the spectra of six bright pulsars have been found~\cite{Majumdar_2018,Pallathadka:2020vwu}.

These signals have been extensively searched for in the spectrum of various targets also at TeV energies as well as with X-rays (see \cite{Batkovi__2021}, \cite{Reynolds:2019uqt} and references therein), each sensitive to different ranges in the ALPs mass range, thus putting constraints which are complementary to other stellar probes and laboratory experiments. 

The standard statistical techniques used in past works are based on the search of ALP-induced characteristic wiggles on top of a spectrum generated by the main astrophysical processes governing each source, which is expected to produce rather smooth and featureless shapes, such as power-laws or log-parabolas, with or without characteristic cutoffs. 
Most past works have used standard statistical methods such as likelihood ratio tests or reduced chi-square tests for estimating the statistical presence of such irregularities, assuming as the null hypothesis the smooth, modulation-free spectra, see discussion in \cite{Fermi-LAT:2016nkz,Batkovi__2021}. This requires exploring the ALP parameter space according to a grid of values of the photon-ALP interaction strength  $g_{a\gamma}$ and the axion mass $m_a$. This procedure is thus strongly model-dependent, as it requires computing the prediction for each parameter combination and performing an explicit statistical test with the data for each of them. 

We aim at reducing the model dependency of such searches, exploring whether ALP-photon interactions imprint statistical irregularities that can be identified without scanning the full ALP parameter space. On a similar note, the authors of \cite{Kachelriess:2023fta} have recently explored the discrete power spectrum as a potential model-independent estimator of wiggles in TeV \gr spectra coming from ALPs. More recently, supervised machine-learning-based approaches have also been explored to search for ALP-induced spectral features, further illustrating the interest in data-driven strategies for such signals~\cite{Schiavone:2025vrb}.

In what follows, we outline the adopted simulation of the ALPs-modulated spectra and then present the results of the supervised and weakly supervised classification. 
For simplicity, Galactic pulsars are taken as benchmark sources, for which only the Galactic magnetic field has to be taken into account when simulating the oscillation probability. However, the methodology we present is fully general and can be applied to any source type  and to other energy ranges, e.g. the TeV spectra of Galactic and extragalactic sources, provided that a careful investigation of the performance is carried out with dedicated simulations. In this exploratory setup, we do not attempt to model the full astrophysical population realism, but instead treat modulation as a controlled deformation of otherwise standard, smooth pulsar spectra.

\subsection{Simulation and feature representation} 

We focus on Galactic sources, in particular pulsars, and consider the potential modulation of their \gr spectra in the Galactic magnetic field (GMF). It can be shown that the mixing of photons and ALPs in the presence of a magnetic field becomes efficient at a critical energy $E_c$ that depends on the photon-ALPs interaction strength $g_{a\gamma}$, the axion mass $m_a$, and the magnetic field strength $B$ as~\cite{Majumdar_2018},
\begin{equation}
\label{eqn:ecrit}
E_{c}\simeq 2.5~\mathrm{GeV}  \frac{|m_{a}^2-\omega_{pl}^2|}{1\,\mathrm{neV}} \left(\frac{B_{\perp}}{\mu\rm G}\right)^{-1} \left(\frac{g_{a\gamma}}{10^{-11}~\mathrm{GeV}^{-1}}\right)^{-1} \, , 
\end{equation}
where $\omega_{pl}=0.03~\mathrm{neV}~n_e^{1/2}$  is the plasma frequency in a medium with electron density $n_e$ in electrons per $\mathrm{cm}^{-3}$, and $B_{\perp}$ is the transverse magnetic field. In the strong-mixing regime $E_\gamma \gg E_c$, the conversion probability becomes approximately energy-independent. At energies $E_\gamma\sim E_c$, the mixing is energy-dependent and can lead to oscillatory features in the observed spectra. For typical Galactic parameters, the oscillation length for these phenomena can be expressed as in~\cite{Mirizzi_2017}: 
\begin{equation}
\label{eqn:losc}
l_\mathrm{osc}= 32~\mathrm{kpc}\sqrt{1+(E_c/E_\gamma)^2} \left(\frac{B_\perp}{\mu\mathrm{G}}\right)^{-1}
                                \left(\frac{g_{a\gamma}}{10^{-11}~\mathrm{GeV}^{-1}}\right)^{-1} \, .
\end{equation}  
This implies that, for typical GMFs of order $\mu$G, observable effects for Galactic sources can be expected at GeV energies. The GMF has a non-constant, complex structure, and can be modeled as a superposition of a regular and a turbulent component, see \cite{Jaffe:2019iuk} for a recent review. A widely used model for the regular component is the Jansson-Farrar model~\cite{Jansson2012}. This GMF parametrization was tuned to reproduce Faraday rotation measures and the Galactic synchrotron emission map.  
In our calculations, we use the Jansson-Farrar model for the regular GMF, with parameters as given in \cite{Jansson2012}, while we leave to future work the exploration of the more updated benchmark GMF models defined in \cite{Unger:2023lob} or other effects connected to the turbulent component of the GMF \cite{Carenza:2021alz}. 

In general, the energy-dependent oscillation probability of \gr photons traversing the GMF from any position can be calculated numerically for a given source, magnetic field model, and ALP parameters $(m_a, g_{a\gamma})$.  We use the tool \texttt{GammaALPs}~\cite{Meyer_2021} to simulate the energy-dependent oscillation probability needed to modulate the spectra of \gr pulsars. This includes advanced modelling of the conversion probability and of the electron density, as well as an implementation of the  Jansson-Farrar GMF. 

Given the exploratory nature of our investigation,  we rely on a sample of simulated pulsar spectra obtained as described below. We extract pulsar spectra and properties as reported in the third pulsar catalog (3PC) released by the Fermi-LAT collaboration \cite{Fermi-LAT:2023zzt}. Specifically, for each pulsar listed, we extract its position in the sky and the listed distance (see details on the distance computation in \cite{Fermi-LAT:2023zzt}), and the spectral best fit model together with the estimated uncertainty. This is done by taking into account the uncertainties and covariance matrix of the fitted spectral parameters. We then proceed by fixing the number of energy bins to 50 (we tested that our results are stable against this choice, increasing the number to 150) and extract multiple realizations of the measured spectrum for each source by sampling spectral parameters within their reported uncertainties (including covariance) to generate multiple realizations of spectra consistent with the catalog fit. 
We note that finer energy binning is required for this application with regard to the energy-dependent fluxes from the 4FGL catalog, as used in the previous two applications. This is necessary in order to evidence the spectral modification produced by ALP-photon oscillations.
This is done to obtain multiple realizations of possible measured spectra consistent with the reported uncertainties. This sample of sources is used as a starting set for the unmodulated sources. To produce modulated spectra, we start from the spectra extracted as described above, and multiply by the oscillation probability computed using \texttt{GammaALPs} for their distance and position in the sky. We fix the total number of pulsar spectra to 4000 to ensure sufficient statistics for the classification task.

We note that, when searching for ALP-induced modulations within real observed sources, other effects should be carefully taken into account. Above all, the energy resolution of the observed fluxes as a function of energy guides the choice of the energy binning, and the uncertainties in the energy reconstruction (usually modeled by an energy dispersion matrix) determine the magnitude of the modulation that could be detected within standard searches~\cite{Fermi-LAT:2016nkz}. Since we do not aim here at constraining the ALPs parameter space, we refrain from including these complications, which can be explored in future work. 

\begin{figure}
    \centering
    \includegraphics[width = 0.95\textwidth]{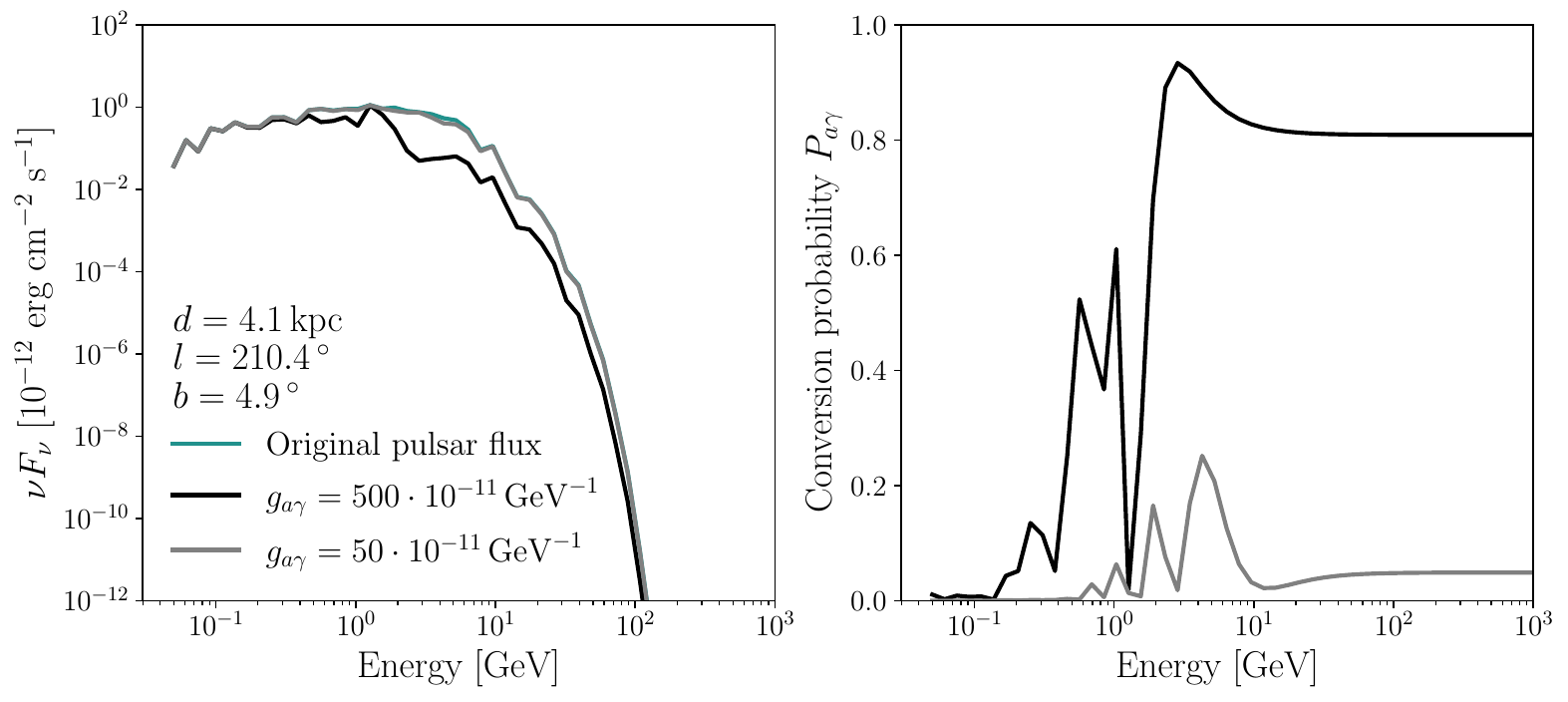}
    \caption{\textit{Left:} Modulated spectrum of a simulated pulsar for two different ALP-photon coupling strengths. The modulation is clearly visible in the spectrum for a stronger coupling. \textit{Right:} Axion-photon conversion probability as a function of energy corresponding to the ALP model and pulsar position shown in the left panel.}
    \label{fig:Sim_PSR_mod_example}
\end{figure}

An example of a modulated spectrum is shown in figure \ref{fig:Sim_PSR_mod_example} for two different ALP-photon coupling strengths of $50 \times 10^{-11}$ GeV$^{-1}$  (gray line) and $500 \times 10^{-11}$ GeV$^{-1}$ (black line) and a sampled pulsar at distance of 4.1~kpc and located in the sky at $l=210.4$ and $b=4.9$ degrees (left panel). We also show (right panel) the computed  conversion probability as a function of energy for the ALP models  and at the pulsar position. The modulation of the ''noisy'' pulsar spectrum is in particular visible as oscillatory features at low energies and as a suppression of the flux at high energies with respect to the original pulsar flux (green line). As expected, the modulation is more significant assuming a stronger coupling. We refer to Appendix~\ref{sec:app_ALPchoice} for a discussion of the ALP parameter choices.

\subsection{Supervised benchmark}

We use the same supervised setup as in the previous section and evaluate the classifier using 5-fold cross-validation. The only modification is an increase in the number of estimators to 400, which yields a modest but consistent improvement in validation performance. The larger ensemble allows for a more flexible exploitation of the higher-dimensional feature space introduced by the increased number of input bins.

To set a benchmark for weakly supervised classification, we first train a supervised classifier on simulated modulated and unmodulated pulsar spectra. We train on 4000 samples, where the modulation due to ALP conversions is applied to a subset determined by the chosen signal fraction. We consider two setups. First, we assume that the modulated pulsar spectra are observable in 10~\% of the pulsar data set, i.e. we consider a signal fraction of 10~\%. We study this case to compare a strongly imbalanced supervised setup (10~\% signal) with the weakly supervised configuration, where the imbalance appears between background and mixed samples rather than between labelled signal and background examples, but where the underlying signal fraction in the mixed sample is also of order 10~\%.

We present the results of the supervised classification in table~\ref{tab:supervised_ALP_9010} for two different configurations of the axion-photon coupling strength $g_{a\gamma}$ and different numbers of energy bins in the pulsar spectra. For the ALP mass, we choose $m_a = 1$~neV. We discuss this choice briefly in appendix~\ref{sec:app_ALPchoice}.

We find that the accuracy is generally high ($>90$~\%), driven by the very low false positive rate. This is a consequence of the class imbalance, as in such classification tasks the classifier tends to favor the majority class in order to minimize overall error. As a result, the true positive rate is low, especially for the case of $g_{a\gamma} = 50 \times 10^{-11}\, \mathrm{GeV}^{-1}$. This is expected, as the modulated spectra are very similar to the unmodulated spectra, and the distinction between the two classes is very challenging. We do not observe a significant difference in classifier performance when increasing the number of energy bins.

\begin{table}[]
    \centering
    \begin{tabular}{l|c|c|c|c}
        \toprule
         $g_{a\gamma} / (10^{-11}\, \mathrm{GeV}^{-1})$ & 50 & 50 & 500 & 500 \\ 
         Number of energy bins & 50 & 150 & 50 & 150 \\ \hline
         Accuracy & 0.910 & 0.907 & 0.953 & 0.953 \\
         TPR & 0.157 & 0.108 & 0.549 & 0.552 \\
         FPR & 0.006 & 0.004 & 0.002 & 0.002 \\ 
         \bottomrule
    \end{tabular}
    \caption{Results from a supervised classification of modulated / not modulated pulsar spectra, assuming an observable modulation in 10~\% of pulsar spectra. We test different configurations of the axion-photon coupling strength $g_{a\gamma}$ and the number of energy bins of the pulsar spectra. Due to the class imbalance, the accuracy is high, but the true positive rate is low.}
    \label{tab:supervised_ALP_9010}
\end{table}

We also test the supervised classification for the case of a signal fraction of 30~\%, i.e. we assume that the modulated pulsar spectra are observable in 30~\% of the pulsar data set. This approximately matches the effective class imbalance encountered in the weakly supervised setup between background and mixed samples. The results are shown in table~\ref{tab:supervised_ALP_7030}. We find that the accuracy is lower than in the 10~\% signal fraction case, but the true positive rate is significantly higher. This is expected, as the class imbalance is less severe and the classifier has many more modulated samples to learn from.

\begin{table}[]
    \centering
    \begin{tabular}{l|c|c|c|c} 
        \toprule
         $g_{a\gamma} / (10^{-11}\, \mathrm{GeV}^{-1})$ & 50 & 50 & 500 & 500 \\
         Number of energy bins & 50 & 150 & 50 & 150 \\ \hline
         Accuracy & 0.775 & 0.793 & 0.911 & 0.921 \\
         TPR & 0.363 & 0.395 & 0.735 & 0.761 \\
         FPR & 0.049 & 0.037 & 0.014 & 0.011 \\
         \bottomrule
    \end{tabular}
    \caption{The same as table~\ref{tab:supervised_ALP_9010}, but for an observable modulation in 30~\% of pulsar spectra. This corresponds to the class imbalance in the weakly supervised setup between the background and mixed classes.}
    \label{tab:supervised_ALP_7030}
\end{table}

These supervised benchmarks establish the dependence of performance on both signal strength and class imbalance, and serve as reference points for the weakly supervised results discussed below.

\subsection{Weakly supervised results}

We now apply the weakly supervised  setup to the simulated pulsar spectra in order to identify modulated spectra among predominantly unmodulated ones. This reflects the realistic expectation that ALP-induced modulations, if present, would only affect a subset of sources depending on their sky position and magnetic-field environment. We do not select pulsars according to their expected modulation strength; instead, we modulate randomly chosen spectra and assess whether the weakly supervised approach can recover them statistically.

We construct a dataset of 4000 spectra and split it into a background sample $B$ (70~\%) and a mixed sample $M$ (30~\%). Within the mixed sample, only a fraction $f$ of spectra is modulated. For $f=10~\%$, this corresponds to an effective signal fraction of $0.3 \times 0.1 = 3~\%$ in the full dataset. Thus, the task is intrinsically challenging, as the overwhelming majority of spectra are unmodulated.

Table~\ref{tab:CWoLa_ALP} summarises the weakly supervised performance for $f=10~\%$ and a 70/30 background-to-mixture split. As in the supervised 10~\% setup, the overall accuracy remains high ($\sim 0.9$), primarily because the dataset is dominated by background-like spectra. Compared to both supervised benchmarks, however, the weakly supervised classifier exhibits a higher false positive rate and a reduced true positive rate, reflecting the fact that the BvM objective optimises separation between $B$ and $M$, rather than directly between modulated and unmodulated spectra.

\begin{figure}
    \centering
    \includegraphics[width = 0.75\textwidth]{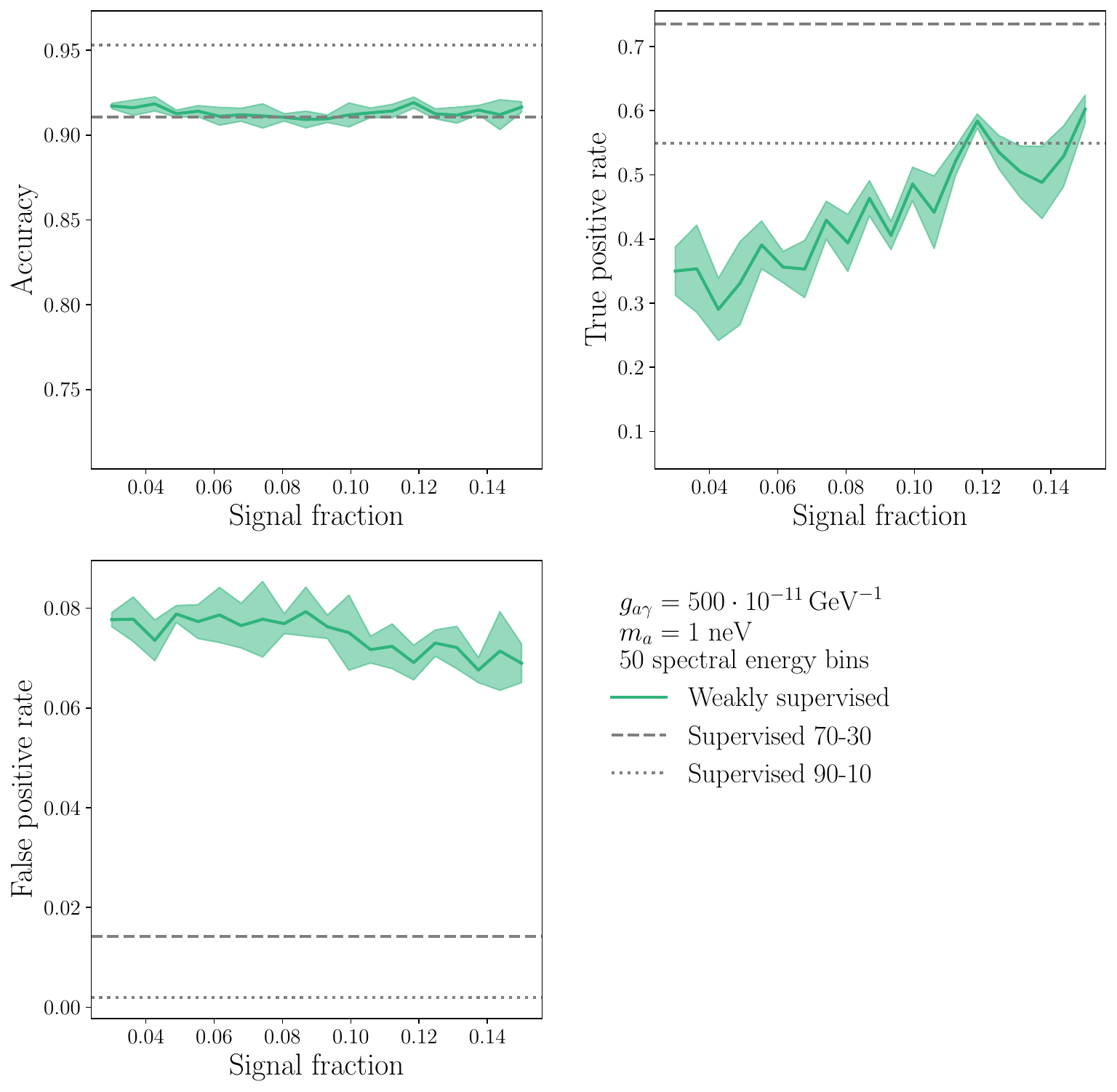}
    \caption{Weakly supervised results for the setup with $g_{a\gamma} = 500 \times 10^{-11}\, \mathrm{GeV}^{-1}$ and 50 pulsar energy bins. We show the supervised classification benchmarks for both the 10~\% and 30~\% signal fractions, see also tables~\ref{tab:supervised_ALP_9010} and~\ref{tab:supervised_ALP_7030}.}
    \label{fig:alp_g500_nE50} 
\end{figure}

\begin{figure}
    \centering
    \includegraphics[width = 0.75\textwidth]{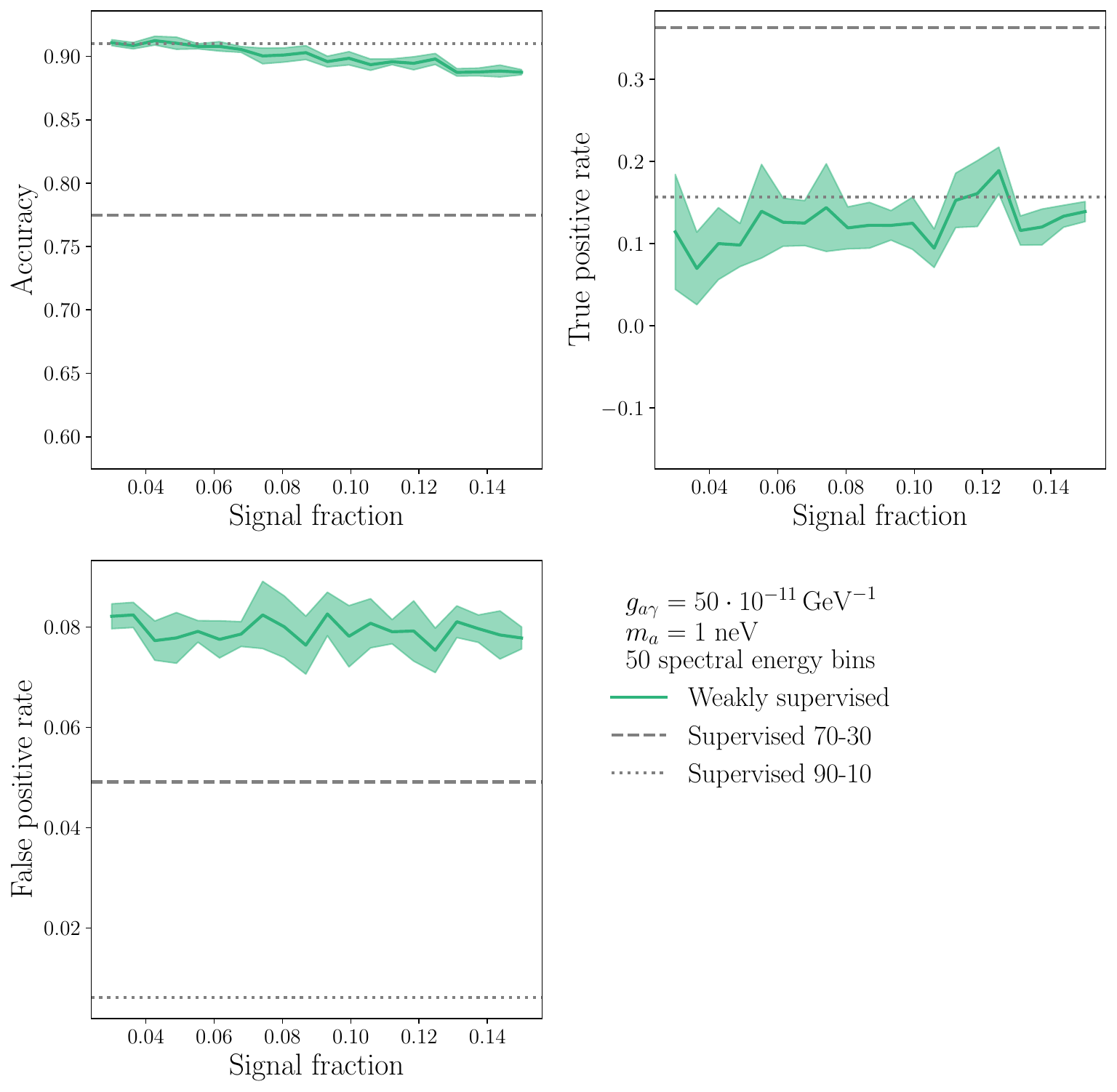}
    \caption{Same as figure~\ref{fig:alp_g500_nE50}, but for $g_{a\gamma} = 50 \times 10^{-11}\, \mathrm{GeV}^{-1}$.}
    \label{fig:alp_g50_nE50}
\end{figure}

For 50 energy bins, the dependence on signal fraction is shown in figures~\ref{fig:alp_g500_nE50} and~\ref{fig:alp_g50_nE50} for strong and weak coupling, respectively. Results are averaged over five independent runs.

In the strong-coupling case ($g_{a\gamma}=500 \times 10^{-11}\,\mathrm{GeV}^{-1}$), the accuracy approaches that of the supervised 30~\% benchmark at larger signal fractions. However, the true positive rate remains systematically lower and the false positive rate higher. This behaviour is expected: although the training imbalance (70/30) resembles the supervised 30~\% case, the effective signal fraction in the weakly supervised setup is substantially smaller (only $0.3 \times f$). Consequently, the decision boundary learned from $B$ versus $M$ remains conservative with respect to true signal identification.

As the signal fraction increases, the true positive rate improves and can approach supervised values for sufficiently large $f$, but this occurs together with an elevated false positive rate. Given the small number of modulated spectra relative to the full dataset, even modest background contamination significantly impacts the false positive rate.

In the weak-coupling case ($g_{a\gamma}=50 \times 10^{-11}\,\mathrm{GeV}^{-1}$), both supervised and weakly supervised approaches struggle to distinguish modulated from unmodulated spectra. The modulation amplitude is comparable to the simulated spectral noise, and the classifier predominantly predicts the majority class. We verified that this limitation is not driven by specific source positions or distances; rather, it reflects the intrinsic difficulty of detecting such weak modulations under realistic measurement uncertainties.

Overall, the weakly supervised approach is capable of highlighting ALP-like spectral distortions in the strong-coupling regime without requiring labelled signal examples during training. However, in contrast to the dark-matter subhalo case, the performance degradation relative to supervised learning is more pronounced, reflecting the subtlety of the spectral features and the very small effective signal fraction.

In future work, we will explore alternative regions of the ALP parameter space, updated Galactic magnetic-field models, and refined simulation setups. Extensions toward simulation-based inference methods may provide a complementary, less model-dependent framework for the identification of spectral irregularities \cite{Bhattacharjee:2025ofd}.

\begin{table}[]
    \centering
    \begin{tabular}{l|c|c|c|c}
        \toprule
         $g_{a\gamma} / (10^{-11}\, \mathrm{GeV}^{-1})$ & 50 & 50 & 500 & 500 \\ 
         Number of energy bins & 50 & 150 & 50 & 150 \\ \hline
         Accuracy & 0.899 & 0.898 & 0.912 & 0.915 \\
         TPR & 0.125 & 0.106 & 0.486 & 0.439 \\
         FPR & 0.078 & 0.078 & 0.075 &  0.071 \\ 
         \bottomrule
    \end{tabular}
    \caption{Excerpt of the weakly supervised results at a signal fraction of 10~\%. We train with a 70/30 class imbalance between the background and mixed classes.}
    \label{tab:CWoLa_ALP}
\end{table}

% --------------------------------------------------
% CONCLUSIONS
% --------------------------------------------------

\section{Conclusion}\label{sec:conclusion}

In this work we have explored the potential of weakly supervised classification for less model-dependent searches of new phenomena in $\gamma$-ray spectra. Rather than relying on fully labelled signal examples, we considered a background--versus--mixture (BvM) setup in which only mixed samples with an unknown signal fraction are available during training. We studied three representative scenarios: the separation of pulsars and AGN as a controlled benchmark, a search for dark-matter subhalos, and spectral irregularities induced by axion-photon oscillations.

Across these case studies, weak supervision consistently recovers a non-trivial fraction of the performance of fully supervised classifiers, despite not having access to labelled signal samples during training. In the pulsar-AGN benchmark, where the classes are well separated, weak supervision approaches the supervised baseline. For dark-matter subhalos, the method remains competitive over a range of signal fractions, albeit with a systematic trade-off between true positive rate and false positive rate that reflects the structure of the BvM setup. The ALP-induced modulation scenario represents a more challenging task, where the signal can be subtle and easily obscured by statistical fluctuations; here both supervised and weakly supervised approaches encounter intrinsic limitations when the modulation amplitude falls below the level of spectral noise.

A central observation is that the performance of weak supervision depends not only on the intrinsic separability of signal and background, but also on the signal fraction and the composition of the mixed samples used during training. In the background--versus--mixture framework, small signal fractions naturally lead to more conservative classification boundaries, suppressing false positives at the cost of reduced signal efficiency. Increasing the effective influence of the mixed sample enhances the sensitivity to signal-like events, but also admits a larger fraction of background. This characteristic trade-off appears consistently across the different physics scenarios considered in this work.

From a physics perspective, weakly supervised methods offer a complementary strategy to traditional, strongly model-dependent searches. In particular, they can highlight anomalous or signal-like subsets of data without committing to a specific signal template during training. Such candidate subsets may then be studied with dedicated likelihood analyses, multiwavelength observations, or follow-up measurements. While weak supervision cannot replace model-dependent inference once precise parameter constraints are required, it may provide a flexible discovery-oriented framework for prioritizing potentially interesting sources, especially in situations where signal modelling is uncertain or multiple new-physics hypotheses are conceivable.

Several directions for future work emerge from our study. A natural next step is the application of the BvM framework directly to real Fermi-LAT data, carefully controlling selection effects and validating the same-background assumption. On the methodological side, the exploration of alternative architectures and representation-learning strategies may improve sensitivity to subtle spectral features. Finally, combining weak supervision with simulation-based inference techniques could provide a path toward less model-dependent searches that nevertheless allow for quantitative parameter constraints once candidate signals are identified.

Overall, our results indicate that weakly supervised classification constitutes a viable addition to the toolbox for indirect searches for new physics in high-energy astrophysical data.

%==========================================================
% ACKNOWLEDGMENTS
%==========================================================
\acknowledgments
We thank Alexander M\"uck for discussions and valuable comments on the manuscript. MK acknowledges support by the Deutsche Forschungsgemeinschaft (DFG, German Research Foundation) under grant 396021762 -- TRR 257 ``Particle Physics Phenomenology after the Higgs Discovery''. Simulations and neural network training were performed using computing resources granted by RWTH Aachen University under project \texttt{rwth0754}.
SM acknowledges the support of the French Agence Nationale de la Recherche (ANR) under the grant ANR-24-CPJ1-0121-01, and support of the European Union's Horizon Europe research and innovation program for support under the Marie Sklodowska-Curie Action HE MSCA PF-2021,  grant agreement No.10106280, project \textit{VerSi}.

% --------------------------------------------------
% APPENDIX
% --------------------------------------------------

\appendix

\section{Generation of additional $\gamma$-ray spectra with normalizing flows}
\label{sec:app_NF}

The 4FGL catalog contains a limited number of classified sources per class. 
To study anomaly-detection performance in controlled but realistically imbalanced settings, 
we generate additional $\gamma$-ray spectra that follow the observed AGN distribution. 
This allows us to construct background samples with the same size as the full catalog 
while retaining control over class composition. 
In particular, we consider the case of searching for pulsar-like anomalies within an AGN-dominated catalog.

To generate additional spectra, we employ a normalizing flow. 
Normalizing flows are generative models that transform a simple base distribution 
into a complex target distribution through a sequence of invertible mappings. 
They allow exact likelihood evaluation and efficient sampling, 
which makes them well suited for density estimation in moderately low-dimensional data. 
In our case, the dimensionality is set by the number of energy bins (seven in the baseline setup), 
so the data space remains sufficiently compact for stable flow training.

We use a normalizing flow with spline-based coupling layers~\cite{Durkan2019NeuralSF,Durkan2019CubicSplineF}, which has proven suitable for learning multimodal distributions. The flow is trained on the preprocessed AGN spectra from the 4FGL catalog. 
As in the classification task, we apply a logarithmic transformation to the fluxes 
and standardize all inputs to zero mean and unit variance. 
The architecture consists of eight sequential transformations, 
each composed of two coupling layers. 
The main hyperparameters are summarised in table~\ref{tab:cusp_hyperparameters}. 
The model is trained using maximum likelihood for 1000 epochs with a batch size of 128.

\begin{table}[t]
    \def\arraystretch{1.15}
    \caption{Hyperparameter settings for the spline-based normalizing flow.}
    \centering
    \begin{tabular}{ll}
    \toprule
    Number of knots $K$ & 10 \\
    Number of transformations & 8 \\
    Layer structure in each coupling block & $[32, 32]$ \\
    Activation function & ReLU \\
    Learning rate & $10^{-4}$ \\
    Batch size & 128 \\
    \bottomrule
    \end{tabular}
    \label{tab:cusp_hyperparameters}
\end{table}

\begin{figure}[t]
	\center
	\includegraphics[width=.65\textwidth]{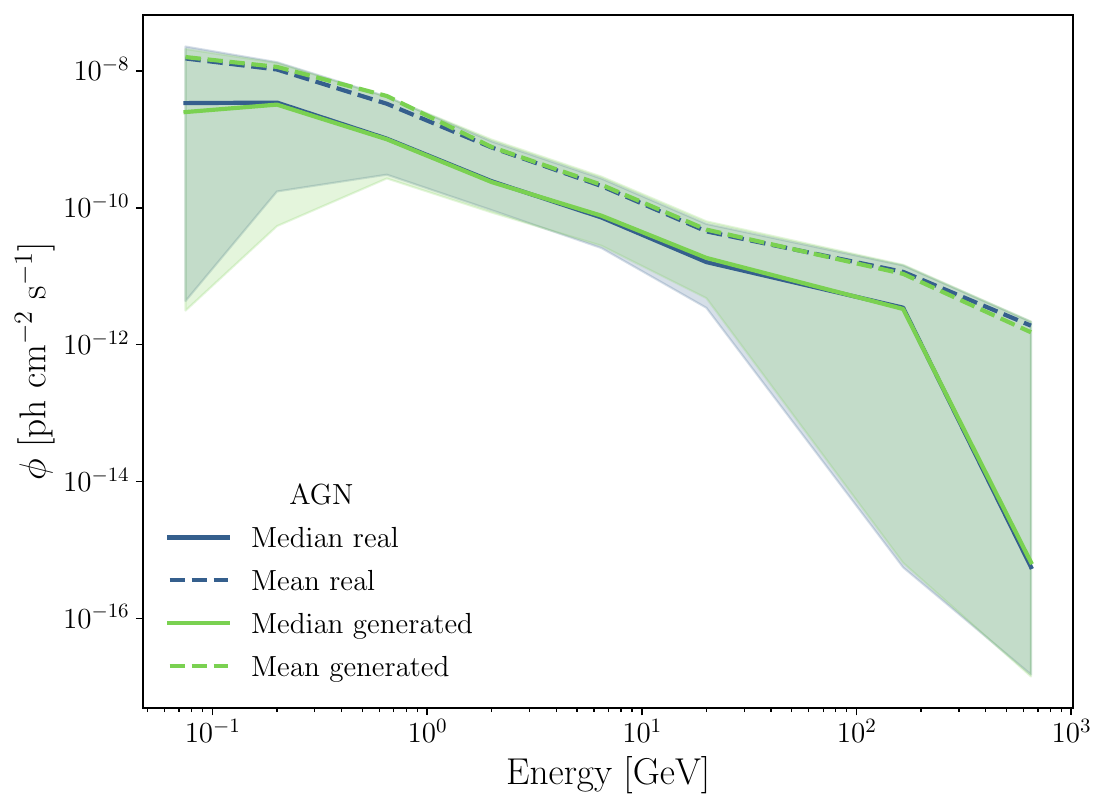}
		\caption{Flux distribution of generated (green) and observed (blue) AGN spectra.
                The shaded band indicates the 68\% interval, the solid line the median,
                and the dashed line the mean.}
		\label{fig:AGN_gen_spectra}
\end{figure}

To validate the generative performance, we compare generated and observed spectra. 
Figure~\ref{fig:AGN_gen_spectra} shows the flux distribution of generated AGN spectra 
together with the real 4FGL AGN spectra, matching the total number of sources. 
The median, mean, and 68\% percentile bands agree well across all energy bins. 

\begin{figure}[t]
	\center
	\includegraphics[width=.85\textwidth]{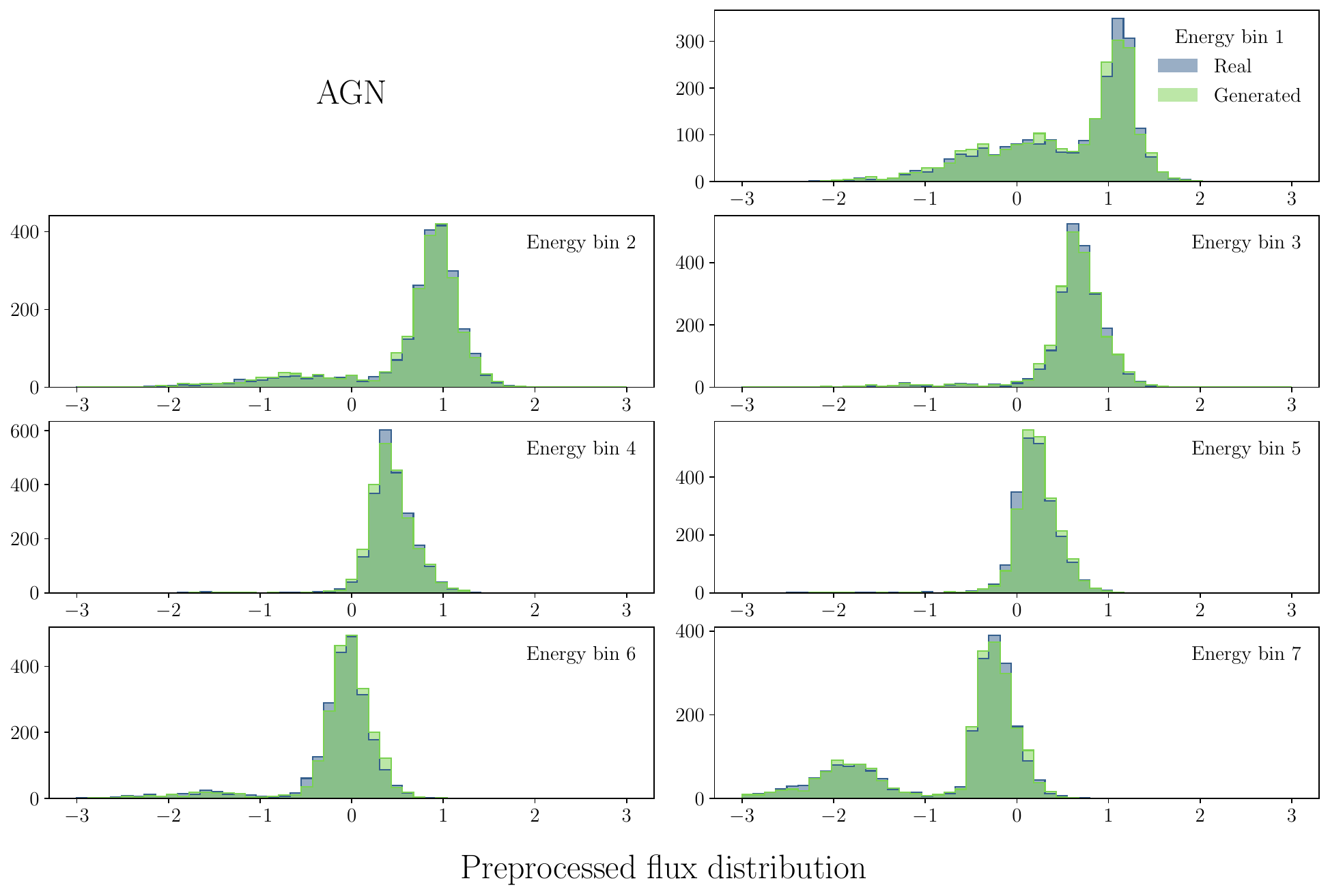}
		\caption{Distributions of the preprocessed flux values per energy bin for generated (green)
and observed (blue) AGN spectra.}
		\label{fig:AGN_gen_ebins}
\end{figure}

A more differential comparison is shown in figure~\ref{fig:AGN_gen_ebins}, 
where we display the distributions of the preprocessed flux values per energy bin. 
The generated spectra reproduce both the width and the shape of the observed distributions. 
In particular, bimodal structures visible in some bins are well captured, 
indicating that the flow successfully learns the relevant substructure of the AGN population.

As an additional quantitative test, we train a classifier to distinguish real from generated spectra. 
The best-performing network achieves an accuracy of $52.6\%$ and an AUC of $0.53$, 
consistent with random guessing within statistical fluctuations. 
The classifier is trained on an 80/20 train-test split, and early stopping is applied. 
The near-random discrimination performance indicates that the generated spectra 
do not contain obvious artifacts that would allow a systematic separation from real data.

We have repeated this procedure for different subsets of the 4FGL catalog 
(AGN subclasses, pulsars, and the full set of classified sources) 
and consistently observe similar generative quality. 
In the applications presented in the main text, 
the normalizing flow is used exclusively for controlled data augmentation 
and not as a standalone anomaly score.

\section{ALP parameter choices}
\label{sec:app_ALPchoice}

The choice of the ALP parameter space, represented by $(m_a, g_{a\gamma})$, 
is motivated by the exploratory nature of our study. 
We focus on the modulation of pulsar spectra in the Fermi-LAT energy band 
in a parameter region currently probed in the literature on ALP-induced spectral modulations, 
namely masses of order $10^{-9}$~eV and photon couplings down to $\sim 10^{-11}\,\mathrm{GeV}^{-1}$.

\begin{figure}[t]
    \centering
    \includegraphics[width = 0.95\textwidth]{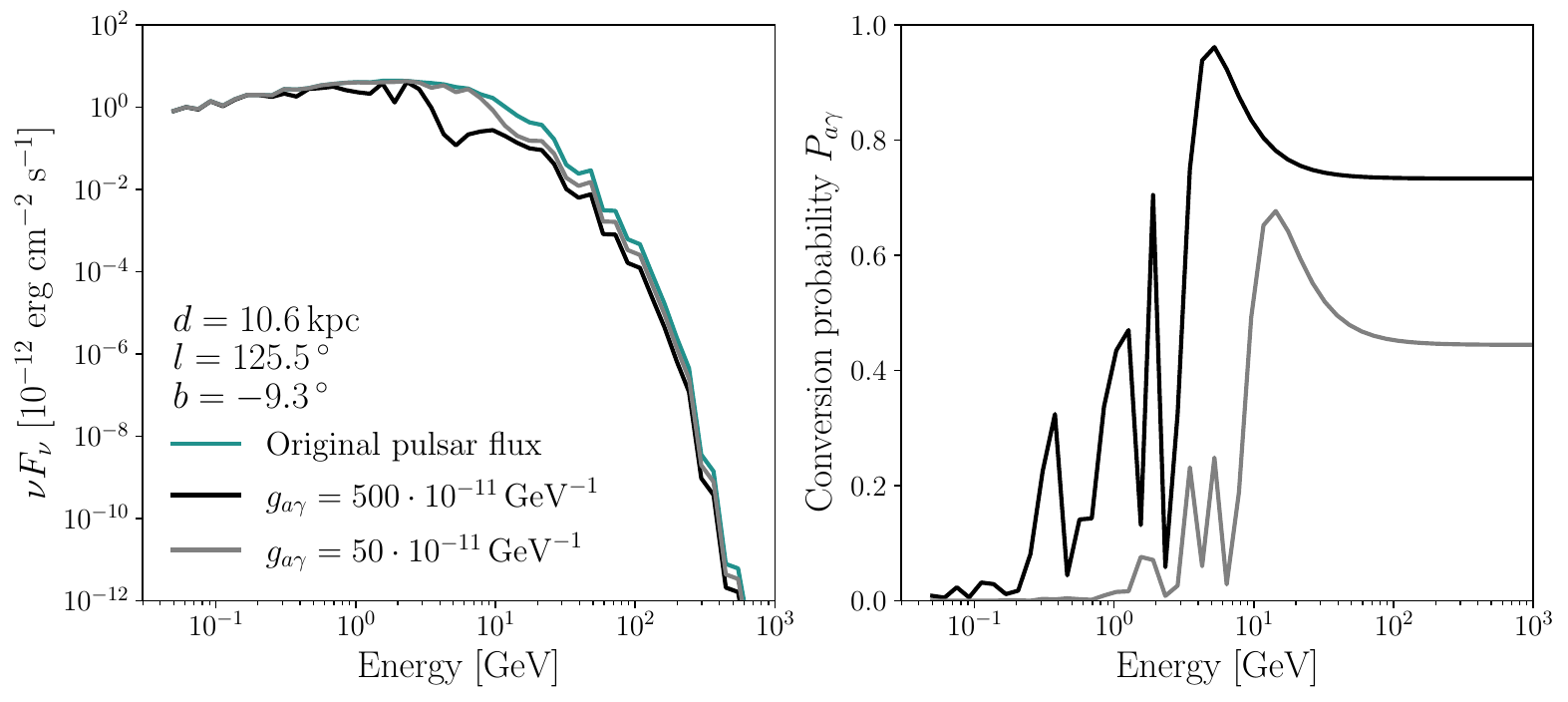}
    \caption{Example of an ALP-induced modulation for a different sampled source
position and intrinsic pulsar spectrum, using the same benchmark
parameters as in section~\ref{sec:alps}.
Left: intrinsic (blue) and modulated (orange) $\gamma$-ray spectrum.
Right: corresponding photon survival probability as a function of energy.}
    \label{fig:Sim_PSR_mod_example_2}
\end{figure}

For the benchmark points used in section~\ref{sec:alps}, 
the chosen parameters lie in a region that is tested by current constraints. 
and lead to oscillatory conversion probabilities in the Fermi-LAT energy band, 
as illustrated in figures~\ref{fig:Sim_PSR_mod_example} 
and~\ref{fig:Sim_PSR_mod_example_2}.

\begin{figure}[t]
    \centering
    \includegraphics[width = 0.95\textwidth]{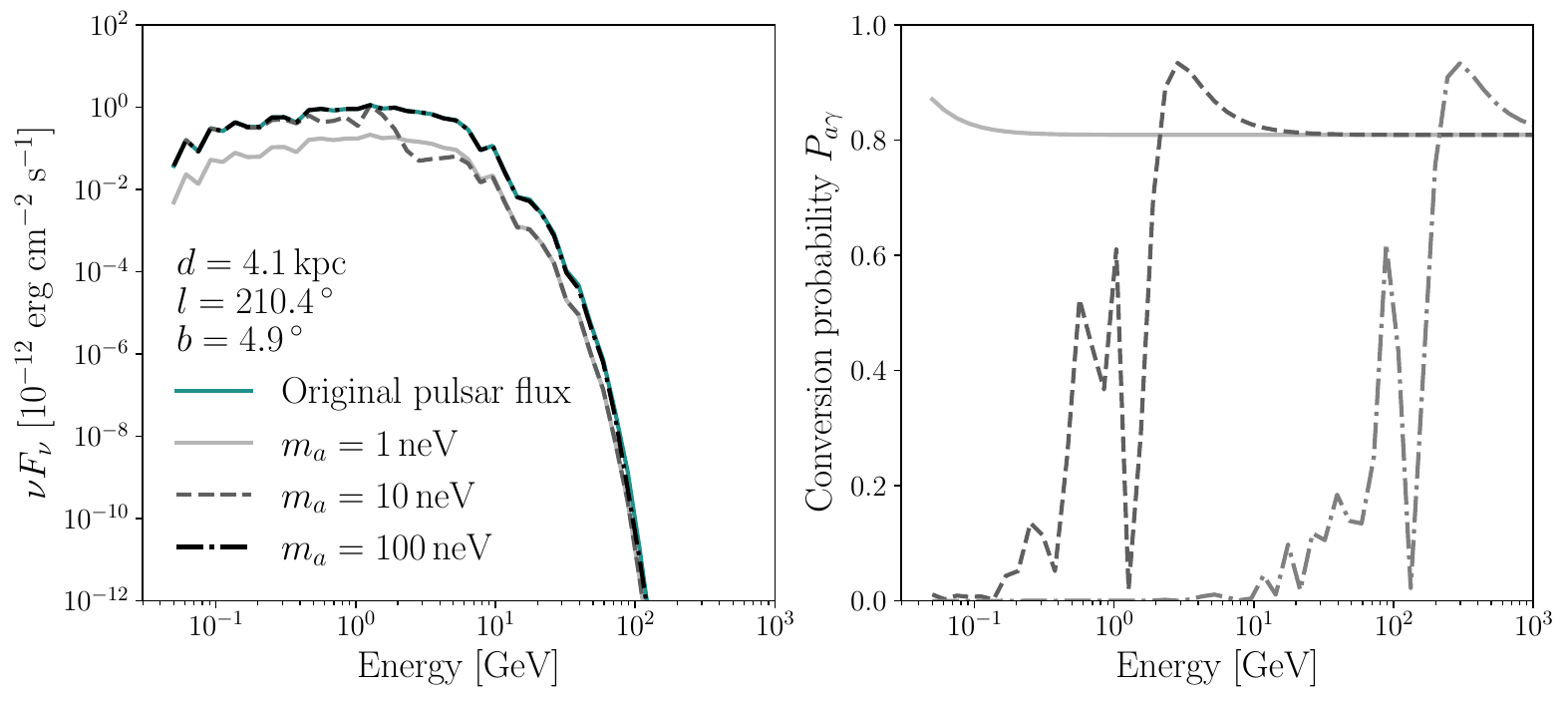}
    \caption{Effect of varying the ALP mass on the spectral modulation,
keeping the photon coupling fixed.
Left: intrinsic (blue) and modulated (orange) $\gamma$-ray spectrum.
Right: photon survival probability illustrating the shift
of the oscillation pattern within the Fermi-LAT energy band.}
    \label{fig:Sim_PSR_mod_example_mass}
\end{figure}

Different choices of the axion mass shift the characteristic oscillation pattern 
towards higher or lower energies, as shown in figure~\ref{fig:Sim_PSR_mod_example_mass}.

% --------------------------------------------------
% BIBLIOGRAPHY
% --------------------------------------------------

\bibliographystyle{JHEP}
\bibliography{bib}

\end{document}